\input epsf
\newcommand{\be}{\begin{equation}}
\newcommand{\ee}{\end{equation}}
\newcommand{\bea}{\begin{eqnarray}}
\newcommand{\eea}{\end{eqnarray}}
\newcommand{\bean}{\begin{eqnarray*}}

\newcommand{\eean}{\end{eqnarray*}}
\font\upright=cmu10 scaled\magstep1
\font\sans=cmss10
\newcommand{\ssf}{\sans}
\newcommand{\stroke}{\vrule height8pt width0.4pt depth-0.1pt}
\newcommand{\Z}{\hbox{\upright\rlap{\ssf Z}\kern 2.7pt {\ssf Z}}}

\newcommand{\C}{{\rlap{\rlap{C}\kern 3.8pt\stroke}\phantom{C}}}
\newcommand{\R}{\hbox{\upright\rlap{I}\kern 1.7pt R}}
\newcommand{\CP}{\C{\upright\rlap{I}\kern 1.5pt P}}
\newcommand{\PP}{\hbox{\upright\rlap{I}\kern 1.5pt P}}

\newcommand{\identity}{{\upright\rlap{1}\kern 2.0pt 1}}

\newcommand{\taubf}{\mbox{\boldmath $\tau$}}
\newcommand{\bn}{{\bf n}}

\documentstyle[12pt,a4wide]{article}

\begin{document}
\title{\vskip -70pt
\begin{flushright}
{\normalsize UKC/IMS/98-33} \\
{\normalsize DAMTP-1998-109} \\
\end{flushright}\vskip 50pt
{\bf \LARGE \bf To be or knot to be?}} 
\author{Richard A. Battye$^{\ \dagger}$ and Paul M. Sutcliffe$^{\ \ddagger}$\\[10pt]
\\{\normalsize $\dagger${\sl Department of Applied Mathematics and Theoretical Physics,}}
\\{\normalsize{\sl University of Cambridge,}}
\\{\normalsize {\sl Silver Street, Cambridge, CB3 9EW, U.K.}}
\\{\normalsize {\sl Email : R.A.Battye@damtp.cam.ac.uk}}\\ 
\\{\normalsize $\ddagger$ {\sl Institute of Mathematics, University of Kent at Canterbury,}}\\
{\normalsize {\sl Canterbury, CT2 7NZ, U.K.}}\\
{\normalsize{\sl Email : P.M.Sutcliffe@ukc.ac.uk}}\\}
\date{}
\maketitle
\newpage
{\bf It has been suggested recently that knots might exist as stable soliton solutions
 in a simple three-dimensional classical field theory~\cite{FN}, opening up a wide range
 of possible applications in physics and beyond. 
We have re-examined and extended this work in some detail using a combination of analytic approximations
 and sophisticated numerical algorithms. Although most of the assertions made 
in the earlier work of Faddeev and Niemi~\cite{FN} as to the 
structure of solitonic solutions are incorrect, the basic idea that complex soliton
 configurations exist in this model is sound. For charges between one and eight, 
we find solutions which exhibit a rich and spectacular variety of phenomena, including
 stable toroidal solitons with twists, linked loops and also knots. The physical process 
which allows for this variety is the reconnection of string-like segments.}

\smallskip 

The specific model under consideration is an $O(3)$ variant of the Skyrme
 model~\cite{Sk} with Lagrangian density in (3+1)-dimensions given by~\cite{Fa}
\be
{\cal L}=\partial_\mu \bn\cdot \partial^\mu \bn
-\frac{1}{2}(\partial_\mu \bn \times \partial_\nu \bn)\cdot
(\partial^\mu \bn \times \partial^\nu \bn)\,,
\label{lag}
\ee
where the field $\bn=(n_1,n_2,n_3)$ takes values on the 2-sphere, that is, ${\bf n}^2=1$. 
The two parts to the Lagrangian are known as the sigma model and Skyrme terms respectively; 
the latter being included to stabilize the solitons against radial scaling as in Derricks
 theorem~\cite{derrick}. 

In order for a solution to have finite energy, the field must be fixed at spatial
 infinity, say $\bn_{\infty}=(0,0,1)$, and hence the domain is compactified from $\R^3$
 to $S^3$, so that at any fixed time we have a map 
$\bn: S^3 \mapsto S^2.$ Therefore, each field configuration is characterized by a topological
 charge $Q$, since $\pi_3(S^2)=\Z$,
 which is known as the Hopf invariant. This can be defined formally
 in terms of the pullback of the area two-form $\omega$ on the target $S^2$. If $F=\bn^*\omega$ is
 the pullback under $\bn$ onto the domain, then it must be exact, $F=dA$, since the second cohomology
 group of the 3-sphere is trivial. One can then construct the Hopf charge by integrating 
the Chern-Simons term over $\R^3$,
\be 
Q={1\over  4\pi^2}\int d^3x \,F\wedge A\,.
\ee
It can also be interpreted heuristically as the linking number between field lines, in
 contrast to other topological characteristics, such as baryon number in the Skyrme model,
 which are generally winding numbers. Since the preimage of a point on the target $S^2$ is 
a closed loop in $\R^3$, the preimages of any two distinct points will be linked exactly 
$Q$ times. Just as in the case of other solitonic models, there exists a bound on the energy
 $E$ of a configuration with charge $Q$~\cite{KR}. However, the dependence is not linear
but rather $E\ge c|Q|^{3/4}$ and $c=16\pi^2 3^{3/8}\approx 238$.

Clearly, the equations of motion for this model are not analytically tractable and hence 
a numerical approach is required. We have used a code originally designed to investigate 
Skyrmions in the $O(4)$ version of the model~\cite{BS1,BS2,BS3} in three spatial dimensions 
on a discretized Cartesian grid,  which can be trivially modified for the situation 
under consideration here. It was run on a Silicon Graphics Origin 2000 parallel super-computer
 which now has 44 fast R10000 processors and 20Gb of memory. We used spatial discretizations
 with $100^3$ points and for the lower charges convergence was achieved in around 50 
CPU hours, but the higher charge configurations took very much longer. 

The basic minimization procedure is to create suitably random initial conditions with a 
 specific Hopf charge, which are then evolved under the full equations of motion. The 
originally static initial conditions turn potential energy into kinetic energy as 
they evolve, and this kinetic energy is periodically removed when the potential energy 
begins to increase. A more detailed exposition of this method for locating minima in 
these models is given in ref.~\cite{BS3}. The only technically difficult aspect is to 
produce initial conditions with a given Hopf charge which are devoid of any symmetries
 and are, hence, suitably random. This can be done in analogy to an approach used to 
construct Skyrme configurations.

If $U({\bf x})$ is a smooth $SU(2)$ field with winding
 number $B$ between two 3-spheres, which correspond to the compactified $\R^3$ and the 
group manifold of $SU(2)$, then writing the components of $U$ in terms of complex numbers 
$Z_0$ and $Z_1$ as 
\be
U=\left(\begin{array}{cc}Z_0&-\bar Z_1\\
Z_1&\bar Z_0\end{array}\right)\,,
\ee
where $|Z_0|^2+|Z_1|^2=1$, one can construct an $O(3)$ field from 
$\bn=Z^\dagger\taubf Z\,,$
where $\tau_j$ are the Pauli matrices and the column vector is $Z=(Z_0,Z_1)^T$. 
It is easy to see that the vector defined above has unit length and satisfies
 the boundary condition $\bn(\infty)=\bn_\infty$ if the original $SU(2)$ field is the 
identity at spatial infinity. Furthermore, it can be shown that $Q=B$~\cite{Me} and so the 
problem is reduced to finding some appropriate Skyrme field configurations.

One can construct suitable Skyrme fields using the rational map approach~\cite{HMS}. 
In spherical polar coordinates $(r,\theta,\phi)$, this involves specifying the angular 
distribution of the field using a rational function $R(z)$ with degree $B$ of the complex
 variable $z=e^{i\phi}\tan(\theta/2)$, and a radial profile function $f(r)$, which
 satisfies the boundary conditions $f(0)=\pi$ and $f(\infty)=0$. A particularly simple 
example is $R(z)=z^B$, which for $B=1$ corresponds to a spherically symmetric hedgehog 
Skyrmion, and for $B>1$ describes an axially symmetric toroidal field. Using 
the arguments above this allows us to construct an  $O(3)$ field with Hopf charge $Q=B$ which
 has only axial symmetry, since the Hopf projection breaks any spherical symmetry. Of course this
 is not particularly helpful from the point of view of the minimization procedure, but a 
simple modification to include non symmetric wiggles can remove all the symmetries and hence 
provide suitably random initial conditions. The precise map that we use for this is 
\be
R(z)=z^{Q}\left[1+a\cos\left({m\phi^2\over 2\pi}\right)\right]\,,
\ee
which corresponds to a torus with $m\in\Z$ non symmetric wiggles of amplitude $a\in[0,1)$.

We have relaxed the solutions for $Q=1$ to 8 and we have displayed our results by plotting
 several interesting quantities in fig.~1. The first is the preimage of the vector
 $(0,0,-1)$, which defines the position of the soliton. In reality it is difficult to 
compute the locus of this preimage in the discretized domain and hence we will plot 
isosurfaces of the vector $(0,0,-1+\epsilon)$, where $\epsilon\approx 0.2$ is small. This
 allows us to easily visualize the solitons position as the core of a tube rather than a 
single line. We will also explicitly display the linking number, thereby verifying the Hopf
 charge, by plotting in a similar way to the position the loci of two independent points 
$(0,-1,0)$ and $(0,0,-1)$. Finally, we also plot the isosurfaces of energy density. A more 
detailed discussion of the structure of each of the solitons is given in ref.~\cite{BS4}.

For both $Q=1$ and 2 we see that the non-symmetric initial conditions relax quickly back to
 the axially symmetric configurations which are well described by appropriate radial profile functions and 
the rational maps $R(z)=z$ and $R(z)=z^2$ for $Q=1$ and $Q=2$ respectively. In both cases 
the position of the soliton is a circle, but for $Q=1$ the energy density is an axially 
symmetric lump which is localized at the origin, while for $Q=2$ it is localized on a torus 
just inside the locus of the position. Upon examination of the linking structure, one can
 confirm the Hopf charge; the $Q=1$ solution being linked once and the $Q=2$ linked 
twice. 

These two cases had previously been studied in refs.~\cite{FN,GH} using an axially 
symmetric approach which reduces the problem effectively to two dimensions. We agree with
 both groups as to the structure of the $Q=2$ solution, but in the case of $Q=1$, where they
 disagree, we find agreement with ref.~\cite{GH}. Since our approach is completely general,
 using random initial conditions on a three dimensional Cartesian grid with no symmetries, it is
 difficult to believe that our results could be misleading. Therefore, this suggests that 
the numerical approach used in ref.~\cite{FN} is flawed.

\medskip
\begin{center}
\begin{tabular}{|c|c|c|}
 \hline 
$Q$ & Soliton Energy & Torus Energy\\
\hline 
1 & 504  &  505  \\
2 & 835  &  836  \\
3 & 1157 &  1181 \\
4 & 1486 &  1542 \\
5 & 1808 &  1974 \\
6 & 1981 &  2361 \\
7 & 2210 &  2600 \\
8 & 2447 &  3050 \\
\hline
\end{tabular}
\end{center}
{\bf Table 1 } : Energy of the relaxed soliton and torus solutions for charges one to eight.
Note that the difference in energies of the soliton and torus solutions for the first
two charges is a reflection of the accuracy of our  numerical computation of the energy.
\medskip

We have also computed the total energy inside the discretized grid. Although there are some small 
systematic errors in making an absolute measurement of the total energy of the soliton since the grid
 is finite (see ref.~\cite{BS4} for a detailed discussion), the relative values for different charges
 should yield important qualitative information. In particular, we find that our values  for the energy
 of the $Q=1$ and $Q=2$ solitons, tabulated in table 1, are consistent with those of ref.~\cite{GH}

For $Q=3,4$ and 5 we find that the solutions are not axially symmetric and hence could not 
have been found using the methods employed in refs.~\cite{FN,GH}. In each case the locus of
 the position sweeps out a closed loop which is twisted, the twists not being related to 
those in the initial conditions. The energy density isosurfaces are now not toroidal, being 
more reminiscent of pretzels with 2,3 and 4 holes respectively, but each one is also twisted 
to fit inside the locus of the position. Once again the Hopf charge can be verified by 
reference to the linking structure. We have also investigated toroidal configurations for 
these values of $Q$. As one might expect, if we use symmetric initial conditions the symmetry 
is preserved by the relaxation process and hence we can make an estimate of the energies of
 these configurations. As shown in table 1, we find that energies of these symmetric saddle
 point solutions are larger than those for the minima which have twists. We should note that
 we found no evidence for the existence of a stable trefoil configuration for $Q=3$, as 
suggested in ref.~\cite{FN}. We will discuss later why we believe that it is unlikely that
 such a configuration will exist.

Above $Q=5$ there appears to be a dramatic change in the structure of the soliton solutions. 
At $Q=6$ we see that the position of the soliton is no longer a single connected loop, but
 consists of two linked disjoint loops. The fact that the position itself has linking number
 one, makes the counting of the Hopf charge $Q$ more subtle. In fact a careful examination 
of the linking structure reveals that this configuration resembles two linked $Q=2$ solitons.
 The Hopf charge is not simply additive in the case of a link since when a field line passes
 through the intersection it should be counted twice. Note that the energy density isosurface
does not have the form of two linked loops, but is concentrated in the region inside the 
position loops. For $Q=7$ and $Q=8$ even more exotic solutions are produced. It appears
 that the position of the $Q=7$ soliton has the topology of a trefoil knot, while at $Q=8$
 the solution comprizes of two $Q=2$ solitons doubly linked. It should be emphasized that
 the solutions are relaxed from initial conditions which are perturbed tori and hence the
 crossing --- or reconnection --- of field lines is a requisite. In ref.~\cite{BS4} we discuss
 this process in detail and  we present snapshots of the position during the relaxation 
process.

We should note that in any numerical relaxation procedure one can never be totally sure that
 the global minimum has been found, although as much as is possible has been done --- random 
initial conditions and a fully three-dimensional algorithm. One way to increase confidence 
in numerical results is to establish rules dictating the structure of the solutions, which can
 in the end become predictive. In the case of $O(4)$ Skyrmions we suggested~\cite{BS2,BS3} the
 existence of a pattern in the solutions, which we called Geometric Energy Minimization. In that case we 
had a very precise description of the structure of the solutions because they had point symmetries.
 Here, we have a more qualitative description of the solutions, but we can speculate on some general 
features of an energy minimization principle.

For line-like solitons reducing the length of the soliton will naively reduce the 
energy, but this must be balanced by an increase in gradient energy required to 
impose the correct linking of field lines in a reduced volume. For low charges it
 seems reasonable that the solutions be toroidal, with the twists distributed
 uniformly, but as the charge increases it appears that it is possible to reduce 
the length of string without a great cost in gradient energy by twisting the loop
 so that the links can be packed closer together.
What happens for the higher charges is less well defined, but it is clear that
 having extra links in the position itself can reduce the number of links in each
 of the components, and hence also the length of string required. This is illustrated
 by the $Q=6$ and $Q=8$ solitons, which are composed of two $Q=2$ solitons singly
 and doubly linked respectively. The clear special case  is $Q=7$, the first 
structure where the position is actually knotted. We believe that the reason for 
this is symmetry; it would be impossible to distribute the links symmetrically 
into two components. This suggests that as the charge increases the number of 
linked possibilities will increase rapidly, with string reconnection being the
 mechanism by which one linked structure can metamorphose into another. Clearly,
 more work will be required to construct a more quantitative energy minimization 
principle.

It was suggested in ref.~\cite{FN} that the $Q=3$ soliton had the topology 
of a trefoil knot, but our results clearly suggest that this is not the case.
 One could create such a configuration since it has the correct linking structure,
 but the energy minimization principle that we have discussed above  
explains why this is not likely to be the minimum energy configuration --- the 
length of string required to have a knotted solution with $Q=3$ is too long. 

In conclusion, we have demonstrated numerically that a rich variety of fascinating
 closed, linked and knotted configurations are generic in this model. It is possible
 to explain the nature of these solutions heuristically via a qualitative energy
 minimization principle based on the length of string. The results we have presented
 are very much in keeping with the spirit of ref.~\cite{FN}. However, the details
 are very different, in particular the structure of the $Q=1$ and $Q=3$ solitons.
 This investigation will hopefully act as the basis for work on physical systems 
where this model may be applicable, in particular condensed matter~\cite{Pa} and 
particle physics~\cite{FN2}.
 
\def\jnl#1#2#3#4#5#6{{#1 [#2], {\it #4\/} {\bf #5}, #6. } }
\def\jnltwo#1#2#3#4#5#6#7#8{{#1 [#2], {\it #4\/} {\bf #5}, #6;{\bf #7} #8.} }
\def\prep#1#2#3#4{{#1 [#2], #4.} } 
\def\proc#1#2#3#4#5#6{{#1 [#2], in {\it #4\/}, #5, eds.\ (#6).} }
\def\book#1#2#3#4{{#1 [#2], {\it #3\/} (#4).} }
\def\jnlerr#1#2#3#4#5#6#7#8{{#1 [#2], {\it #4\/} {\bf #5}, #6.
{Erratum:} {\it #4\/} {\bf #7}, #8.} }
\def\prl{Phys.\ Rev.\ Lett.}
\def\pr{Phys.\ Rev.}
\def\pl{Phys.\ Lett.}
\def\np{Nucl.\ Phys.}
\def\prp{Phys.\ Rep.}
\def\rmp{Rev.\ Mod.\ Phys.}
\def\cmp{Comm.\ Math.\ Phys.}
\def\mpl{Mod.\ Phys.\ Lett.}
\def\apj{Ap.\ J.}
\def\apjl{Ap.\ J.\ Lett.}
\def\aap{Astron.\ Ap.}
\def\cqg{Class.\ Quant.\ Grav.} 
\def\grg{Gen.\ Rel.\ Grav.}
\def\mn{M.$\,$N.$\,$R.$\,$A.$\,$S.}
\def\ptp{Prog.\ Theor.\ Phys.}
\def\jetp{Sov.\ Phys.\ JETP}
\def\jetpl{JETP Lett.}
\def\jmp{J.\ Math.\ Phys.}
\def\zpc{Z.\ Phys.\ C}
\def\cupress{Cambridge University Press}
\def\oup{Oxford University Press}
\def\pup{Princeton University Press}
\def\wss{World Scientific, Singapore}

\medskip

\section*{Acknowledgments}

\noindent Many thanks to Ludwig Faddeev, Jens Gladikowski, Meik Hellmund, 
Conor Houghton, Patrick Irwin, Nick Manton, Antti Niemi, Miguel Ortiz, Richard Ward, 
and Wojtek Zakrzewski for useful discussions. We would like to acknowledge the use of 
the SGI Origin 2000 and Power Challenge at DAMTP in Cambridge supported by the HEFCE, SGI,
 PPARC, the  Cambridge Relativity rolling grant and EPSRC Applied Mathematics Initiative
 grant GR/K50641. The work of PMS is supported by  EPSRC grant GR/L88320 and that of RAB by Trinity College. 
We would like to thank Paul Shellard for his tireless efforts to provide 
sufficient computational resources for projects such as this.

\section*{Figure Caption} 

\noindent Fig.~1. For charges $Q=1$ to $Q=8$ we display the following quantities for the 
relaxed soliton solutions: (a) the position of the soliton, (b) the linking number of the 
field lines, (c) an energy density isosurface.

\noindent 


\begin{thebibliography}{99}

\bibitem{FN} 
\jnl{L. Faddeev and A.J. Niemi}{1997}{}{Nature}{387}{58}
\prep{L. Faddeev and A.J. Niemi}{1997}{}{hep-th/9705176}
\bibitem{Sk} 
\jnl{T.R.H. Skyrme}{1962}{}{\np}{31}{556}

\bibitem{Fa} 
\prep{L. Faddeev}{1975}{}{Princeton preprint IAS-75-QS70}

\bibitem{derrick}
\jnl{R. Hobart}{1963}{}{Proc. Roy. Soc. Lond.}{82}{201}\jnl{G. Derrick}{1964}{}{J. Math. Phys.}{5}{1252}

\bibitem{KR}
\jnl{A. Kundu and Y.P. Rubakov}{1982}{}{J. Phys.}{A15}{269}\jnl{A.F. Vakulenko and L.V. Kapitanski}{1979}{}{Dokl. Akad. Nauk USSR}{246}{840}

\bibitem{BS1}
\jnl{R.A. Battye and P.M. Sutcliffe}{1997}{}{\pl}{B391}{150}

\bibitem{BS2}
\jnl{R.A. Battye and P.M. Sutcliffe}{1997}{}{\prl}{79}{363}

\bibitem{BS3}
\prep{R.A. Battye and P.M. Sutcliffe}{1998}{}{DAMTP-1998-108}

\bibitem{Me}
\jnl{U.G. Meissner}{1985}{}{\pl}{154B}{190}

\bibitem{HMS}
\jnl{C.J. Houghton, N.S. Manton and P.M. Sutcliffe}{1998}{}{\np}{B510}{507}

\bibitem{BS4}
\prep{R.A. Battye and P.M. Sutcliffe}{1998}{}{DAMTP-1998-110}

\bibitem{GH}
\jnl{J. Gladikowski and M. Hellmund}{1997}{}{\pr}{D56}{5194}

\bibitem{Pa} 
\proc{N. Papanicolaou}{1993}{}{Singularities in Fluids, 
Plasmas and Optics}{R.E. Caflisch and G.C. Papanicolao}{Kluwer Amsterdam}

\bibitem{FN2}
\prep{L. Faddeev and A.J. Niemi}{1998}{}{hep-th/9807069}

\end{thebibliography}
\end{document}